\documentclass[12pt]{article}
\usepackage{graphicx}
\usepackage{cite}

\newcommand\pubnumber{DPF2015-76}
\newcommand\pubdate{\today}

\def\kennesaw{Department of Physics\\
Kennesaw State University, Kennesaw, GA 30144, USA}
\def\support{\footnote{This material is based upon work supported by the National Science Foundation under Grant No. PHY 1519606.}}

\def\Title#1{\begin{center} {\Large #1 } \end{center}}
\def\Author#1{\begin{center}{ \sc #1} \end{center}}
\def\Address#1{\begin{center}{ \it #1} \end{center}}

\newcommand\pubblock{\rightline{\begin{tabular}{l} \pubnumber\\
         \pubdate  \end{tabular}}}
\newenvironment{Abstract}{\begin{quotation}  }{\end{quotation}}
\newenvironment{Presented}{\begin{quotation} \begin{center} 
             PRESENTED AT\end{center}\bigskip 
      \begin{center}\begin{large}}{\end{large}\end{center} \end{quotation}}

\def\beq{\begin{equation}}
\def\eeq#1{\label{#1}\end{equation}}
\def\eeqn{\end{equation}}

\def\beqa{\begin{eqnarray}}
\def\eeqa#1{\label{#1}\end{eqnarray}}
\def\eeqan{\end{eqnarray}}

\let\bar=\overbar

\def\Dslash{\not{\hbox{\kern-4pt $D$}}}
\def\dslash{\not{\hbox{\kern-2pt $\del$}}}

\def\msb{{\bar{\ssstyle M \kern -1pt S}}}

\def\beq{\begin{equation}}
\def\eeq{\end{equation}}
\def\beqa{\begin{eqnarray}}
\def\eeqa{\end{eqnarray}}

\begin{document}
\begin{titlepage}
\pubblock

\vfill
\Title{High-order threshold corrections for top-pair\\ and single-top production}
\vfill
\Author{Nikolaos Kidonakis\support}
\Address{\kennesaw}
\vfill
\begin{Abstract}
I present results for high-order corrections from threshold resummation to cross sections and differential distributions in top-antitop pair production and in single-top production. I show aN$^3$LO results for the total $t{\bar t}$ cross section as well as for the top-quark transverse-momentum ($p_T$) and rapidity distributions, and the top-quark forward-backward asymmetry in $t{\bar t}$ production. I compare with the most recent Tevatron and LHC data, including at 13 TeV. I also present aNNLO results for cross sections and $p_T$ distributions in $t$-channel, $s$-channel, and $tW$-channel single-top production.
\end{Abstract}
\vfill
\begin{Presented}
DPF 2015\\
The Meeting of the American Physical Society\\
Division of Particles and Fields\\
Ann Arbor, Michigan, August 4--8, 2015\\
\end{Presented}
\vfill
\end{titlepage}
\def\thefootnote{\fnsymbol{footnote}}
\setcounter{footnote}{0}

\section{Introduction}

Top-quark physics has been a centerpiece of high-energy collider experiments, first at the Tevatron and then at the LHC. The correct understanding of the production mechanisms of the top is thus of fundamental importance. Total cross sections and differential distributions of the top quark have been calculated to high theoretical precision and measured by the experiments.

In this talk I present the most recent and most accurate theoretical results for the production of top-antitop pairs, and for single top production. Soft-gluon threshold corrections are added at one order higher than known complete fixed-order results in each case. Thus, I present approximate N$^3$LO (aN$^3$LO) results for $t{\bar t}$ production \cite{NKaNNNLO} based on NNLL resummation \cite{NKtop}. I show that they are in excellent agreement with recent LHC and Tevatron data \cite{tt7lhc,tt8lhc,tt13lhc,tttev,CMS8lhcdiff,CMS13lhcdiff,D0pty,tevafb}. An expansion of the NNLL resummed cross section \cite{NKtop} is used through N$^3$LO in \cite{NKaNNNLO}, thus avoiding the need for a prescription. We prefer fixed-order expansions \cite{NKNNNLO} because the minimal or other prescriptions have chronically produced results that severely underestimate the size of the true cross section, while our fixed-order expansions have been exceedingly accurate in predicting the higher-order corrections. I also present approximate NNLO (aNNLO) results for single-top production \cite{NKstW} which are also in excellent agreement with recent LHC and Tevatron data \cite{tchtev,tchlhc,schlhc,tWlhc,CMS13tch}.  

\begin{figure}[htb]
\centering
\includegraphics[height=2.5in]{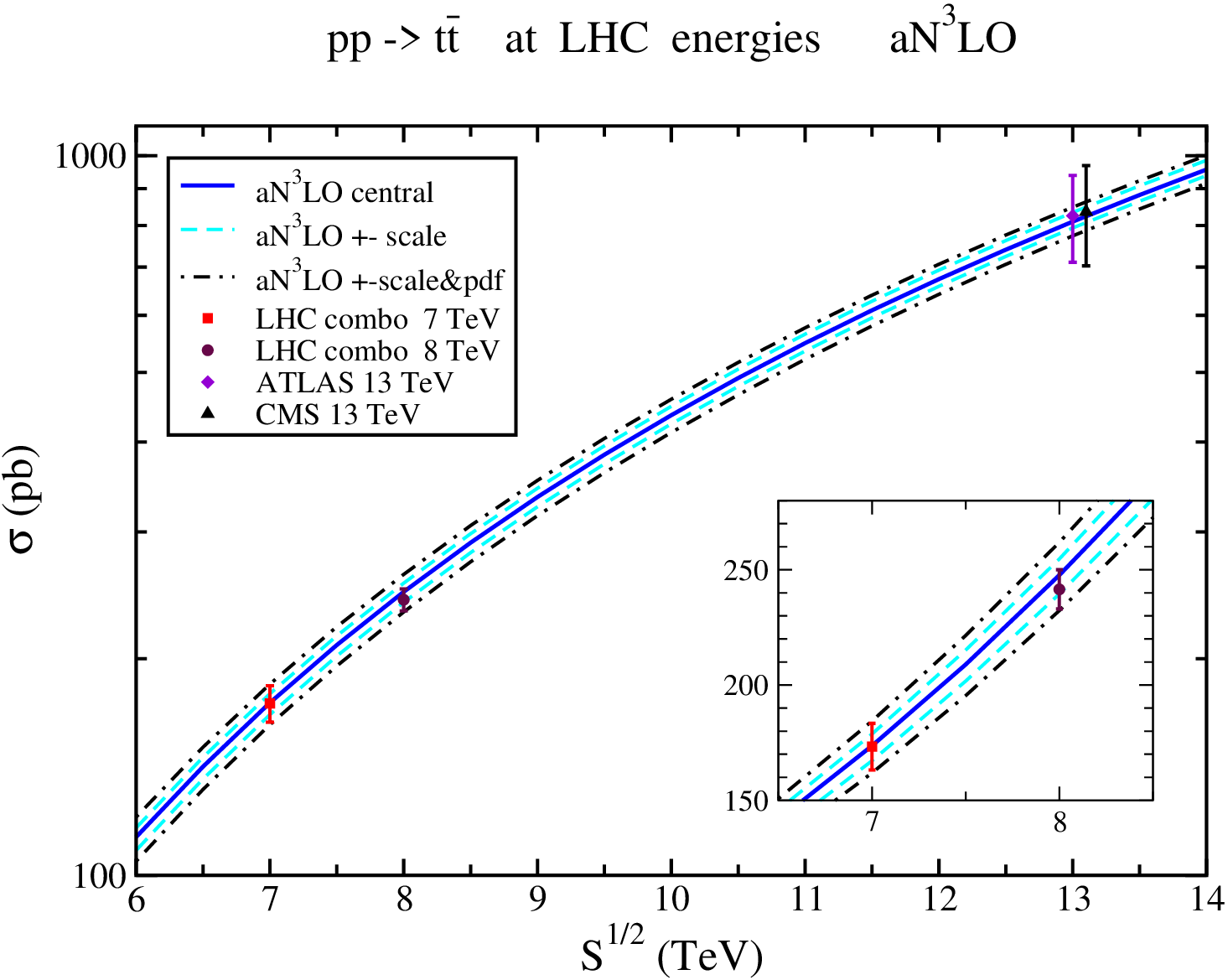}
\caption{Cross sections at aN$^3$LO for $t{\bar t}$ production at the LHC.}
\label{fig:ttlhcplot}
\end{figure}

\section{Top-antitop pair production}

I begin my presentation with aN$^3$LO results for $t{\bar t}$ production. MSTW2008 NNLO pdf \cite{MSTW} are used for all numerical results. I show in Fig. \ref{fig:ttlhcplot} the aN$^3$LO theoretical predictions for the total $t{\bar t}$ cross section as a function of LHC energy. The central result as well as the variations with scale and pdf uncertainties are shown. The theoretical predictions are compared with LHC data from ATLAS and CMS at 7 TeV \cite{tt7lhc} and 8 TeV \cite{tt8lhc}, and with the newest data at 13 TeV \cite{tt13lhc} from Run 2 at the LHC. Remarkable agreement is found for all energies. The recent 13 TeV LHC data have of course still rather large uncertainties, much larger than those of theory. Also excellent agreement is found with 1.96 Tevatron data \cite{tttev} as shown in \cite{NKproc}.

\begin{figure}[htb]
\centering
\includegraphics[height=2.0in]{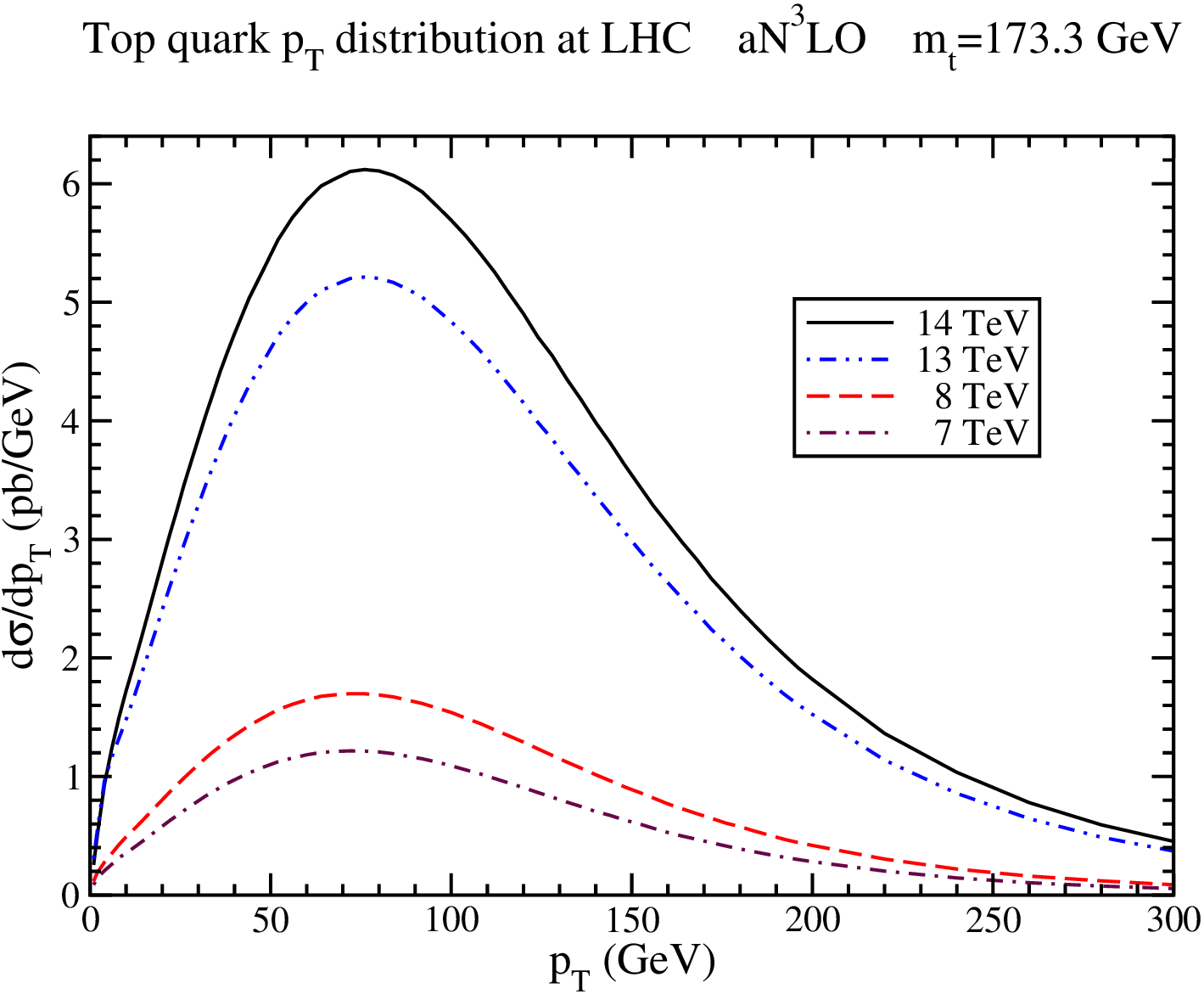}
\hspace{10mm}
\includegraphics[height=2.5in]{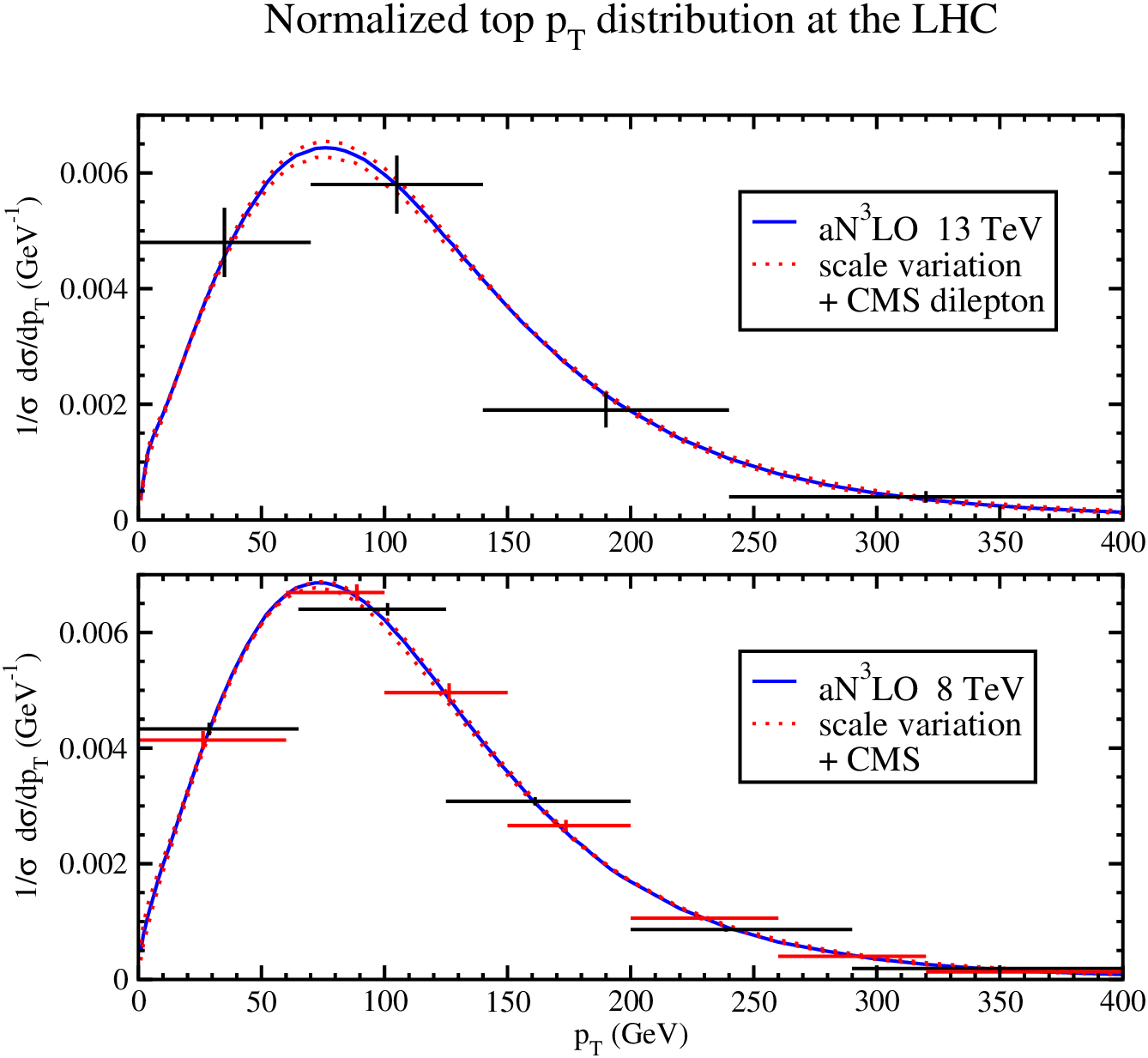}
\caption{Top-quark $p_T$ distributions at aN$^3$LO for LHC energies.}
\label{fig:ptlhcplot}
\end{figure} 

Next I discuss the theoretical results for the $p_T$ distributions. In the left plot of Fig. \ref{fig:ptlhcplot}, I show the central aN$^3$LO results for the top-quark transverse-momentum distributions at 7, 8, 13, and 14 TeV LHC energies. The right plot of Fig. \ref{fig:ptlhcplot} shows the normalized top-quark $p_T$ distributions at 8 TeV (bottom half) compared with CMS data in the dilepton (black) and lepton+jets (red) channels \cite{CMS8lhcdiff}; and at 13 TeV (top half) compared with CMS dilepton data \cite{CMS13lhcdiff}. Excellent agreement is found between our aN$^3$LO results and the data at both energies (and also at 7 TeV as shown in \cite{NKproc}).

\begin{figure}[htb]
\centering
\includegraphics[height=2.0in]{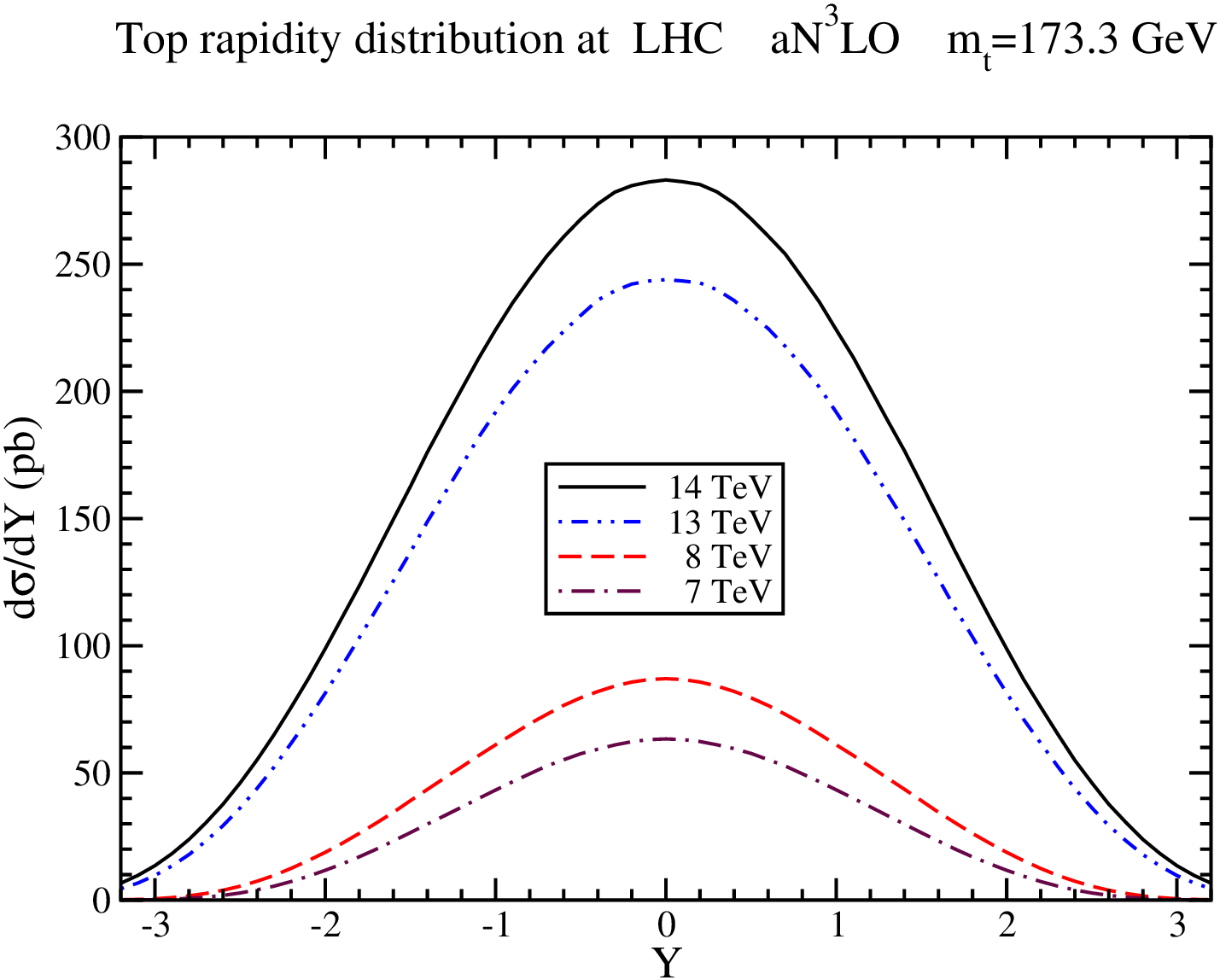}
\hspace{10mm}
\includegraphics[height=2.5in]{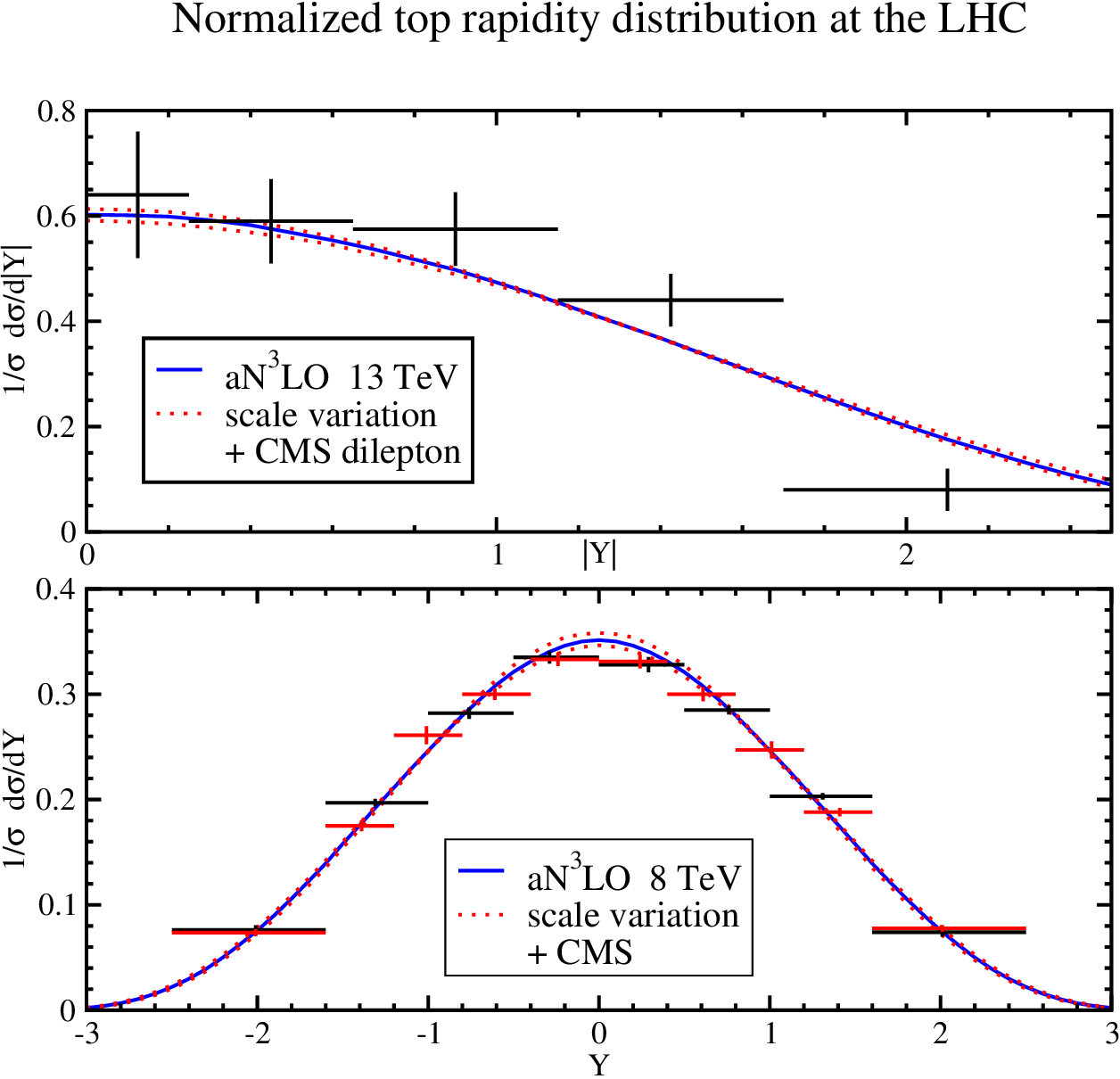}
\caption{Top-quark rapidity distributions  at aN$^3$LO for LHC energies.}
\label{fig:ylhcplot}
\end{figure} 

We continue with theoretical results for the rapidity distributions. In the left plot of Fig. \ref{fig:ylhcplot}, I show the central aN$^3$LO results for the top-quark rapidity distributions at 7, 8, 13, and 14 TeV LHC energies. The right plot of Fig. \ref{fig:ylhcplot} shows the normalized top-quark rapidity distribution at 8 TeV (bottom half) compared with CMS data in the dilepton (black) and lepton+jets (red) channels \cite{CMS8lhcdiff}; and the normalized distribution of the absolute value of the top rapidity at 13 TeV (top half) compared with CMS dilepton data \cite{CMS13lhcdiff}. Excellent agreement of aN$^3$LO theory with data is found at 8 TeV energy (as well as at 7 TeV \cite{NKproc}). At 13 TeV, the error bars of the data are large but there is good overall agreement with theory.

\begin{figure}[htb]
\centering
\includegraphics[height=2.0in]{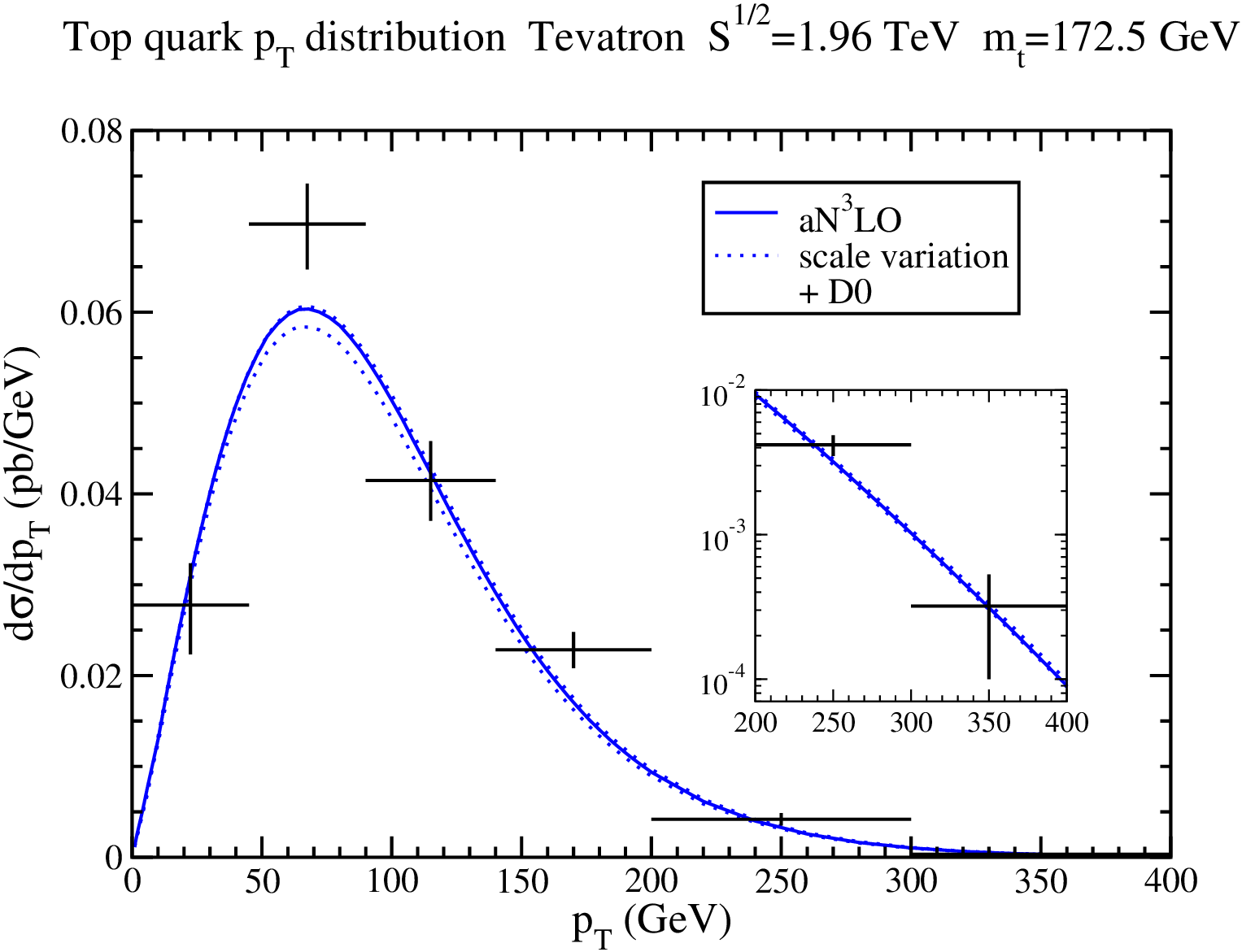}
\hspace{10mm}
\includegraphics[height=2.0in]{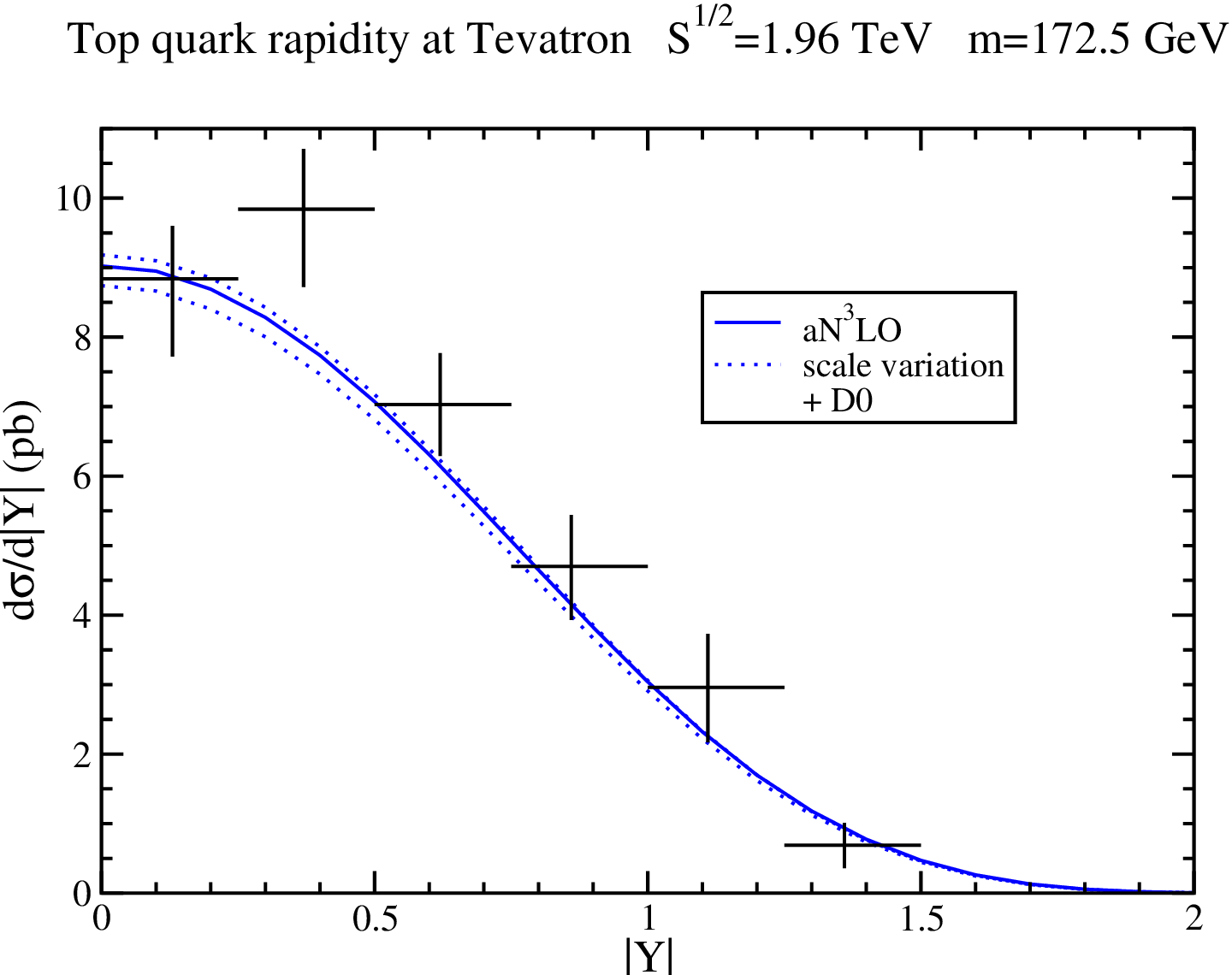}
\caption{Top-quark $p_T$ (left) and rapidity (right) distributions at aN$^3$LO for 1.96 Tevatron energy.}
\label{fig:ptytevplot}
\end{figure} 

In Fig. \ref{fig:ptytevplot}, I present aN$^3$LO results for differential distributions at the Tevatron collider at 1.96 TeV energy. The left plot shows the top-quark $p_T$ distribution, with scale variation, which compares very well with data from D0 \cite{D0pty} (the inset plot shows more clearly the high-$p_T$ region). The right plot of Fig. \ref{fig:ptytevplot} shows the distribution of the absolute value of the top rapidity. Again, the comparison with D0 data \cite{D0pty} shows very good agreement.

The top-quark forward-backward asymmetry, $A_{\rm FB}$, is defined by 
$A_{\rm FB} =[\sigma(Y>0)-\sigma(Y<0)]/\sigma \equiv \Delta\sigma / \sigma$.
Writing explicitly the perturbative series for numerator and denominator, including both electroweak and aN$^3$LO QCD corrections, we can write the expression for $A_{\rm FB}$ as 
\beq
A_{\rm FB}=\frac{\Delta\sigma^{\rm EW}+\alpha_s^3 \, \Delta\sigma^{(1)}
+\alpha_s^4 \, \Delta\sigma^{(2)}+\alpha_s^5 \, \Delta\sigma^{(3)}+\cdots}
{\alpha_s^2 \, \sigma^{(0)}+\alpha_s^3 \, \sigma^{(1)}+\alpha_s^4 \, \sigma^{(2)}
+\alpha_s^5 \, \sigma^{(3)}+\cdots}
\label{AFBnoexp}
\eeq
This expression can be further re-expanded and written as
\beqa
A_{\rm FB} &=& \frac{\Delta\sigma^{\rm EW}}{\alpha_s^2 \, \sigma^{(0)}} 
-\frac{\Delta\sigma^{\rm EW} \sigma^{(1)}}{\alpha_s \, (\sigma^{(0)})^2}
+\frac{\Delta\sigma^{\rm EW}}{(\sigma^{(0)})^3}
\left[(\sigma^{(1)})^2-\sigma^{(0)} \, \sigma^{(2)}\right]
\nonumber \\ && 
{}+\alpha_s \frac{\Delta\sigma^{(1)}}{\sigma^{(0)}}
+\alpha_s^2 \left[\frac{\Delta\sigma^{(2)}}{\sigma^{(0)}}
-\frac{\Delta\sigma^{(1)} \, \sigma^{(1)}}{(\sigma^{(0)})^2}\right]
\nonumber \\ && 
{}+\alpha_s^3 \left[\frac{\Delta\sigma^{(3)}}{\sigma^{(0)}}
-\frac{\Delta \sigma^{(2)} \, \sigma^{(1)}}{(\sigma^{(0)})^2}
+\frac{\Delta \sigma^{(1)} (\sigma^{(1)})^2}{(\sigma^{(0)})^3}
-\frac{\Delta \sigma^{(1)} \, \sigma^{(2)}}{(\sigma^{(0)})^2}\right]+\cdots
\label{AFBexp}
\eeqa

\begin{table}[t]
\begin{center}
\begin{tabular}{|c|c|c|} \hline
aN$^3$LO QCD+EW & $p{\bar p}$ frame & $t{\bar t}$ frame
\\ \hline
$A_{FB}$ \% Eq. (\ref{AFBnoexp}) & $6.4^{+0.5}_{-0.6}$ & $9
.4^{+0.7}_{-0.9}$ \\ \hline
$A_{FB}$ \% Eq. (\ref{AFBexp}) & $6.8 \pm 0.3$ & $10.0 \pm 0.6$ \\ \hline
\end{tabular}
\caption{The top forward-backward asymmetry at 1.96 Tevatron energy.}
\label{tab:AFB}
\end{center}
\end{table}

In Table \ref{tab:AFB}, I show results for the top-quark forward-backward asymmetry at the Tevatron collider at 1.96 TeV energy in both the  $p{\bar p}$ frame and the  $t{\bar t}$ frame. I provide results using both Eq. (\ref{AFBnoexp}) and Eq. (\ref{AFBexp}). The inclusion of the high-order corrections resolves the past disagreement with Tevatron data \cite{tevafb}.

\section{Single-top production}

Next, I discuss aNNLO results for single-top production at the LHC and the Tevatron.

\begin{figure}[htb]
\centering
\includegraphics[height=2.2in]{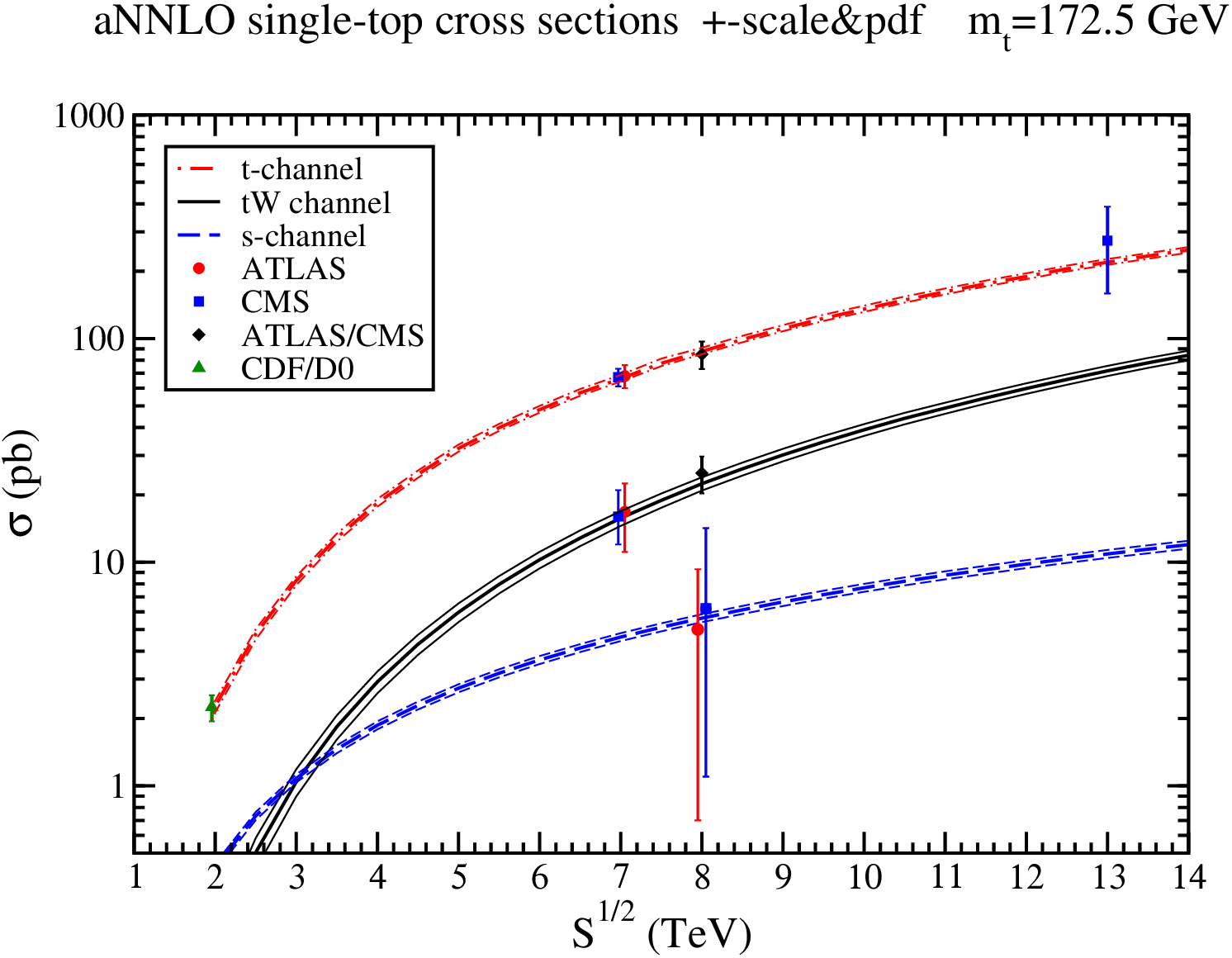}
\hspace{10mm}
\includegraphics[height=2.0in]{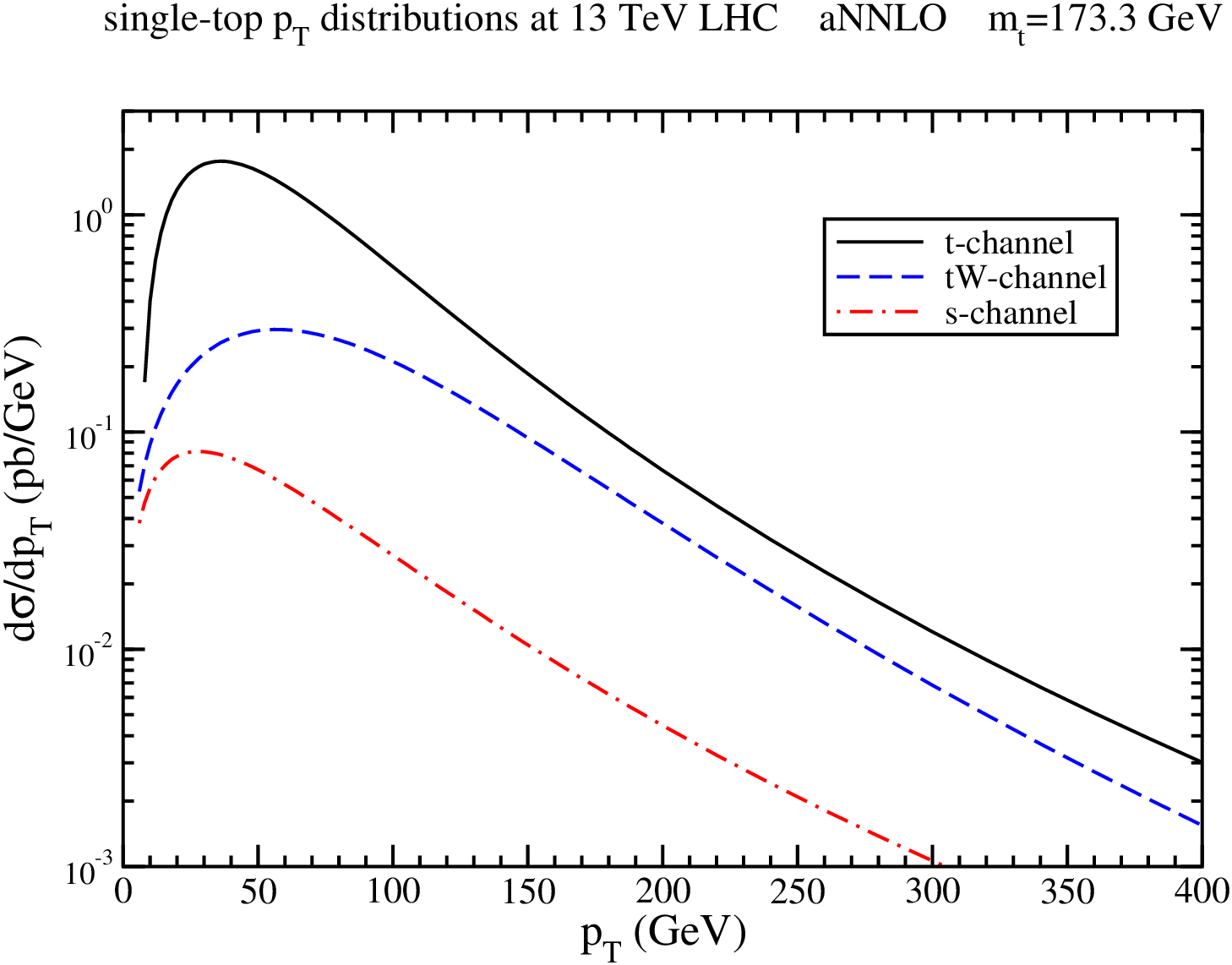}
\caption{Single-top cross sections (left) and $p_T$ distributions (right) at aNNLO for $t$-channel, $s$-channel, and $tW$ production.}
\label{fig:singletopplot}
\end{figure}

In the left plot of Fig. \ref{fig:singletopplot}, I display the aNNLO single-top total cross sections in the $t$-channel, $s$-channel, and $tW$-channel, as functions of collider energy. At 1.96 TeV Tevatron energy, I compare with $t$-channel data from CDF and D0 \cite{tchtev}. At 7 and 8 TeV LHC energies, I compare with $t$-channel data from ATLAS and CMS \cite{tchlhc} and with $tW$-channel data from ATLAS and CMS \cite{tWlhc}. At 8 TeV LHC energy I compare with $s$-channel data from ATLAS and CMS \cite{schlhc}. Finally, at 13 TeV LHC energy I compare with the very recent $t$-channel data from CMS \cite{CMS13tch}. Excellent agreement of theory with data is found for all collider energies.

In the right plot of Fig. \ref{fig:singletopplot} the aNNLO top-quark $p_T$ distributions in single-top production are displayed in all three channels. The $t$-channel distribution is numerically the largest, followed by that for $tW$ production, and last by the $s$-channel distribution.

\section{Conclusions}

I have presented the latest theoretical results for top quark production, both in top-antitop pair and in single-top processes. For $t{\bar t}$ production, I have presented the total cross sections and the top quark differential distributions in $p_T$ and rapidity through aN$^3$LO at LHC and Tevatron energies, as well as the forward-backward asymmetry at the Tevatron. Results for single-top production have been presented through aNNLO for total cross sections and $p_T$ distributions. The high-order corrections are significant, and all theoretical results are in very good agreement with recent analyses from the LHC and the Tevatron.


\begin{thebibliography}{99}

\bibitem{NKaNNNLO} 
N. Kidonakis, Phys. Rev. D {\bf 90}, 014006 (2014) [arXiv:1405.7046 [hep-ph]];
Phys. Rev. D {\bf 91}, 031501(R) (2015) [arXiv:1411.2633 [hep-ph]];
Phys. Rev. D {\bf 91}, 071502(R) (2015) [arXiv:1501.01581 [hep-ph]].

\bibitem{NKtop}
N. Kidonakis, Phys. Rev. Lett. {\bf 102}, 232003 (2009) [arXiv:0903.2561 [hep-ph]]; Phys. Rev. D {\bf 82}, 114030 (2010) [arXiv:1009.4935 [hep-ph]];  
Phys. Rev. D {\bf 84}, 011504 (2011) [arXiv:1105.5167 [hep-ph]].

\bibitem{tt7lhc}
ATLAS and CMS Collaborations, ATLAS-CONF-2012-134, CMS-PAS-TOP-12-003.

\bibitem{tt8lhc}
ATLAS and CMS Collaborations, ATLAS-CONF-2014-054, CMS-PAS-TOP-14-016.

\bibitem{tt13lhc}
ATLAS Collaboration, ATLAS-CONF-2015-033; 
CMS Collaboration, CMS-PAS-TOP-15-005. 

\bibitem{tttev}
CDF and D0 collaborations, Phys. Rev. D {\bf 89}, 072001 (2014) [arXiv:1309.7570[hep-ex]].

\bibitem{CMS8lhcdiff}
CMS Collaboration, arXiv:1505.04480 [hep-ex].

\bibitem{CMS13lhcdiff}
CMS Collaboration, CMS-PAS-TOP-15-010. 

\bibitem{D0pty}
D0 Collaboration, Phys. Rev. D {\bf 90}, 092006 (2014) [arXiv:1401.5785 [hep-ex]].

\bibitem{tevafb}
CDF Collaboration, Phys. Rev. D {\bf 87}, 092002 (2013) [arXiv:1211.1003 [hep-ex]]; D0 Collaboration, Phys. Rev. D {\bf 90}, 072011 (2014) [arXiv:1405.0421 [hep-ex]]. 

\bibitem{NKNNNLO}
N. Kidonakis, Phys. Rev. D {\bf 64}, 014009 (2001) [hep-ph/0010002]; Phys. Rev. D {\bf 73}, 034001 (2006) [hep-ph/0509079].

\bibitem{NKstW}
N. Kidonakis, Phys. Rev. D {\bf 81}, 054028 (2010) [arXiv:1001.5034 [hep-ph]]; 
Phys. Rev. D {\bf 82}, 054018 (2010) [arXiv:1005.4451 [hep-ph]];
Phys. Rev. D {\bf 83}, 091503 (2011) [arXiv:1103.2792 [hep-ph]]; 
Phys. Part. Nucl. {\bf 45}, 714 (2014) [arXiv:1210.7813 [hep-ph]];  
Phys. Rev. D {\bf 88}, 031504 (2013) [arXiv:1306.3592 [hep-ph]].  

\bibitem{tchtev}
CDF and D0 Collaborations, arXiv:1503.05027 [hep-ex].

\bibitem{tchlhc}
ATLAS Collaboration, Phys. Rev. D {\bf 90}, 112006 (2014) [arXiv:1406.7844 [hep-ex]]; CMS Collaboration, JHEP {\bf 12}, 035 (2012)  [arXiv:1209.4533 [hep-ex]]; 
ATLAS and CMS Collaborations, ATLAS-CONF-2013-098, CMS-PAS-TOP-12-002.

\bibitem{tWlhc}
ATLAS Collaboration, Phys. Lett. B {\bf 716}, 142 (2012)  [arXiv:1205.5764 [hep-ex]]; CMS Collaboration, Phys. Rev. Lett. {\bf 110}, 022003 (2013) [arXiv:1209.3489 [hep-ex]]; ATLAS and CMS Collaborations, ATLAS-CONF-2014-052, CMS-PAS-TOP-14-009.

\bibitem{schlhc}
ATLAS Collaboration, Phys. Lett. B {\bf 740}, 118 (2015) [arXiv:1410.0647 [hep-ex]]; CMS Collaboration, CMS-PAS-TOP-13-009.

\bibitem{CMS13tch}
CMS Collaboration, CMS-PAS-TOP-15-004.

\bibitem{MSTW}
A.D. Martin, W.J. Stirling, R.S. Thorne, and G. Watt, 
Eur. Phys. J. C {\bf 63}, 189 (2009) [arXiv:0901.0002 [hep-ph]].

\bibitem{NKproc}
N. Kidonakis, PoS(DIS2015)169 [arXiv:1506.04068 [hep-ph]]; 
PoS(DIS2015)170 [arXiv:1506.04072 [hep-ph]]; 
PoS(RADCOR2015)033 [arXiv:1509.02528 [hep-ph]].

\end{thebibliography}
\end{document}